\documentclass[11pt,english,titlepage]{article}
\usepackage[T1]{fontenc}
\usepackage[latin9]{inputenc}
\usepackage[letterpaper]{geometry}
\usepackage{color}
\usepackage{amsmath}
\usepackage{graphicx}
\usepackage{crossreference}

\makeatletter

\usepackage[round,numbers,compress]{natbib} 

\makeatletter

\usepackage{setspace}

\makeatletter
\renewcommand{\@biblabel}[1]{(#1.)}

\title{Implications of 3-Step Swimming Patterns in Bacterial Chemotaxis}

\author{Tuba Altindal \thanks{ Corresponding author.  Address: 
           Department of Physics and Astronomy, 
           University of Pittsburgh,
   3941 O'Hara Street, Pittsburgh, PA~15260, U.S.A., 
   Tel.:~(412)624-0889, Fax:~(412)624-9163, E-mail:~tua7@pitt.edu }\\
Department of Physics and Astronomy, \\
University of Pittsburgh, Pittsburgh, PA 15260
\and Li Xie \\
Department of Physics and Astronomy, \\
University of Pittsburgh, Pittsburgh, PA 15260
\and Xiao-Lun Wu  \\
Department of Physics and Astronomy, \\
University of Pittsburgh, Pittsburgh, PA 15260 }

\makeatother

\makeatother

\usepackage{babel}

\begin{document}

\title{Implications of 3-Step Swimming Patterns in Bacterial Chemotaxis\newline}
\maketitle
\begin{abstract}
We recently found that marine bacteria \emph{Vibrio alginolyticus}
execute a \emph{cyclic} \emph{3-step} (run-reverse-flick) \textcolor{black}{motility}
pattern that is distinctively different from the 2-step (run-tumble)
pattern of \emph{Escherichia coli}. How this novel swimming pattern
is regulated by cells of \emph{V. alginolyticus} is not currently
known, but its significance for bacterial chemotaxis is self-evident
and will be delineated herein. Using an approach introduced by de
Gennes, we calculated the migration speed of a cell executing the
3-step pattern in a linear chemical gradient, and found that a biphasic
chemotactic response arises naturally. \textcolor{black}{The implication
of such a response for the cells to adapt to ocean environments and
its possible connection to }\textcolor{black}{\emph{E. coli}}\textcolor{black}{'s
response are also discussed.}

\noindent Keywords: \emph{V. alginolyticus, chemotactic behavior,
biphasic response, motility pattern, drift velocity, chemotactic coefficient}
\end{abstract}
\begin{figure}
\includegraphics{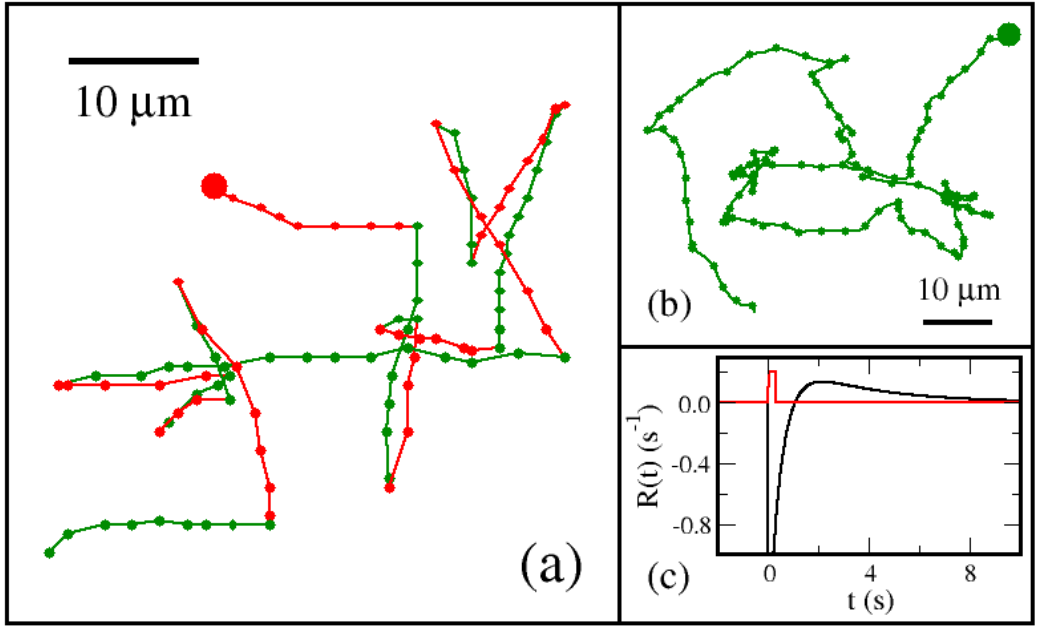}

\caption{\label{fig:Trajectory} Bacterial swimming trajectories of \emph{V.
alginolyticus} (a) and \emph{E. coli} (b). The cells have been selected
amongst many because they are more or less swimming in the focal plane,
50-100 $\mu\text{m}$ above the glass coverslip. The starting points
for both trajectories are indicated by the two large solid dots. The
time lapse between adjacent dots are 0.067 s and 0.13 s for (a) and
(b). The red and green segments in (a) designate the backward and
forward swimming intervals, and transitions from backward to forward
cause flicking, randomizing the swimming direction. Unlike a transition
from forward to backward, which has a directional change $\Delta\theta\simeq\pi$
(or backtracking), a backward to forward transition is random with
$\Delta\theta$ uniformly distributed between 0 and $180\,^{o}$.
(c) A hypothetical response function $R(t)$ of \emph{E. coli} based
on Tu et al.'s model (6) is plotted as the dark curve, and the cell
is stimulated at $t=0$ with a brief pulse of attractant (red curve).
Here, $R(t)=R_{0}\left[\frac{1}{\tau_{m}}\exp\left(-\frac{t}{\tau_{m}}\right)-\frac{1}{\tau_{z}}\exp\left(-\frac{t}{\tau_{z}}\right)\right]$,
where we set $R_{0}=1$, and used the typical \emph{E. coli} methylation
time ($\tau_{m}=3$ s) and phosphorylation time ($\tau_{z}=0.5$ s)
(6, 7).}

\end{figure}

Existing observations made in \emph{Escherichia coli} (\emph{E. coli})
have shown that \textcolor{black}{sensing and motility}\textcolor{red}{{}
}impose different requirements on bacterial chemotactic response (1,
2). The debate on this interesting issue was initiated by the observation
of Block, Segall, and Berg (3) who discovered that the experimentally
measured chemotactic response function $R(t)$ integrated over time
$t$ is zero. In physical terms $R(t)$ can be thought as the Green's
function of the chemotactic network when subjected to an impulsive
or a $\delta$-in-time perturbation. The importance of this null integrated
effect goes without saying, and was immediately recognized by the
investigators as the bacterium\textquoteright{}s means of sensing.
In their words (3), {}``the bacterium compare\textcolor{black}{s}
the information received in the past one second with that received
over the previous three seconds.'' In effect, the double-lobe response
function, \textcolor{black}{which is displayed in Fig. \ref{fig:Trajectory}(c)},
allows the bacterium to react to fast temporal variations of a chemical
signal $c(t)$ but not to its \textcolor{black}{dc}\textcolor{red}{{}
}component, \textcolor{black}{enabling the cell to adapt to a wide
range of chemical concentrations.} Using a macroscopic diffusion argument,
it was suggested by Schnitzer et al. (4) that a finite memory time
is required for a bacterium to migrate in a linear chemical gradient;
without the memory effect (or $R(t)\simeq\delta(t)$), it was concluded
that the chemotactic coefficient $\kappa=V/\nabla c$ or the drift
velocity $V$ would be zero, where $V$ is in the direction of the
chemical gradient $\nabla c$. \textcolor{black}{However, de} Gennes
pointed out that the macroscopic diffusion approach ignored important
correlations between bacterial swimming and the underlying chemical
gradient (5). By taking into account such correlations, de Gennes
showed that the optimal (or a fast) response for migration in a linear
gradient is an exponential function with a decay rate determined by
the cell's \textcolor{black}{memory time $\tau$.} He further pointed
out that the double-lobe response function observed in \emph{E. coli}
could only reduce the migration speed in the gradient. 

Contributing to this stimulating debate is the finding of Clark and
Grant (1), who argued that while a cell needs a fast drift speed in
a concentration gradient, it is equally important for the cell to
localize once the top of the gradient is reached. They showed that
the single-lobe function proposed by de Gennes is inadequate for cell
localization. By imposing the co-requirements of being able to localize
as well as to migrate, they demonstrated that the optimal response
function is biphasic, which is in remarkably good agreement with the
one measured in the experiment (3). This observation led Clark and
Grant to conclude that the biphasic response in \emph{E. coli} perhaps
reflects a compromised need of the cells in\textcolor{red}{{} }\textcolor{black}{different
environments}. \textcolor{black}{A recent study also suggested that
the laboratory observed bacterial response corresponds to the maximin
strategy that ensures the highest minimum uptake of nutrient for any
profile of concentration (2).}

Recently, we found that the swimming pattern of \textcolor{black}{the
}marine bacterium \emph{V. alginolyticus} is a \emph{cyclic 3-step
}\textit{\textcolor{black}{\emph{process}}} (8), where a cell swims
forward for a time interval $\Delta_{f}$ and it then backtracks by
reversing the motor direction for a time $\Delta_{b}$. However, upon
resuming forward swimming, the bacterial flagellum flicks causing
the cell body to veer in a new direction. This type of motility pattern
is very different from that of \textit{E. coli}, which exhibit a run-tumble
pattern. By way of introduction a typical trajectory of \emph{V. alginolyticus}
and of \emph{E. coli} are presented respectively in \textcolor{black}{Fig.
\ref{fig:Trajectory}}(a) and (b). For the \emph{V. alginolyticus'}
trajectory, Fig. \ref{fig:Trajectory}(a), the forward and the backward
segments are designated by green and red, respectively, for clarity.
We termed this novel swimming pattern a \textcolor{black}{run-reverse-flick}\textcolor{green}{{}
}process. The last (flicking) step is functionally equivalent to a
tumble in \emph{E. coli}, allowing the bacterium to randomly select
a direction, and a new cycle ensues. Despite the fact that run and
reverse intervals, $\Delta_{f}$ and $\Delta_{b}$, as well as the
flicking angle $\Delta\theta$ are stochastic, the 3-step cycle is
deterministic and has been observed in different \emph{V. alginolyticus}
strains and in a swimming buffer with and without a chemical gradient
(8). In a steady state without a chemical gradient, we found that
the probability density functions $P(\Delta_{b})$ and $P(\Delta_{f})$
are statistically independent and have long exponential tails (or
a Poissonian-like behavior) with the mean intervals $\tau_{b}\simeq\tau_{f}\simeq0.3\, s$.
\textcolor{black}{However, when a point source of chemoattractant
is present, the cells can quickly migrate along the gradient and form
a tight pack around the source. }

\textcolor{black}{The biochemical network that regulates the activity
of }\textcolor{black}{\emph{E. coli}}\textcolor{black}{{} ~motor is
reasonably well understood (9). While this is not the case for }\textcolor{black}{\emph{V.
alginolyticus}}\textcolor{black}{, it cannot deter our progress because
we know that even for very diverse microorganisms, such as }\textcolor{black}{\emph{E.
coli~}}\textcolor{black}{{} and }\textcolor{black}{\emph{Bacillus subtilis~}}\textcolor{black}{{}
that are roughly one billion years apart according to the recently
constructed phylogenetic tree (10), the fundamental mechanism of regulation
is still similar, i.e. a ligand binding to a receptor triggers a cascade
of chemical reactions. The end product of the reaction is a chemically
modified protein, called the response regulator (CheY-P), that binds
to the motor, causing it either to rotate CCW (}\textcolor{black}{\emph{B.
subtilis}}\textcolor{black}{) or CW (}\textcolor{black}{\emph{E. coli}}\textcolor{black}{).
The basic aim of different microorganisms is also the same, namely
guided by chemical signals, the cell is directed towards the source
of chemoattractant and away from chemorepellent. According to the
phylogenetic tree (10), }\textcolor{black}{\emph{V. alginolyticus}}\textcolor{black}{{}
appears to be much closer to }\textcolor{black}{\emph{E. coli}}\textcolor{black}{{}
than }\textcolor{black}{\emph{B. subtilis}}\textcolor{black}{, suggesting
that there is much in common between these different bacterial species.
Indeed in }\textcolor{black}{\emph{V. alginolyticus}}\textcolor{black}{,
one can identify chemotaxis genes that are largely homologous to }\textcolor{black}{\emph{E.
coli~}}\textcolor{black}{{} with the exception of }\textcolor{black}{\emph{cheV}}\textcolor{black}{{}
that is absent in }\textcolor{black}{\emph{E. coli}}\textcolor{black}{{}
but is present in }\textcolor{black}{\emph{B. subtilis}}\textcolor{black}{.
A recent study moreover showed that the phosphorylated CheY in }\textcolor{black}{\emph{V.
alginolyticus}}\textcolor{black}{{} causes the polar flagellar motor
to reverse the direction from CCW to CW, similar to }\textcolor{black}{\emph{E.
coli}}\textcolor{black}{{} (11). }

\textcolor{black}{It is clear that the 3-step swimming pattern is
significantly different from the well-studied 2-step swimming pattern
of run and tumble, and it has strong implications for bacterial chemotaxis,
which can be characterized by an effective diffusion coefficient $D$
and a drift velocity $V$ in the presence or absence of a chemical
gradient. The calculation below illustrates that cells executing the
3-step swimming pattern can exhibit rich chemotactic behaviors, and
the variations can be acted on by natural selection so that a particular
response emerges. Below we will illustrate these new aspects of bacterial
chemotaxis based on our findings of the 3-step process.}

\begin{figure}
\includegraphics[height=16cm]{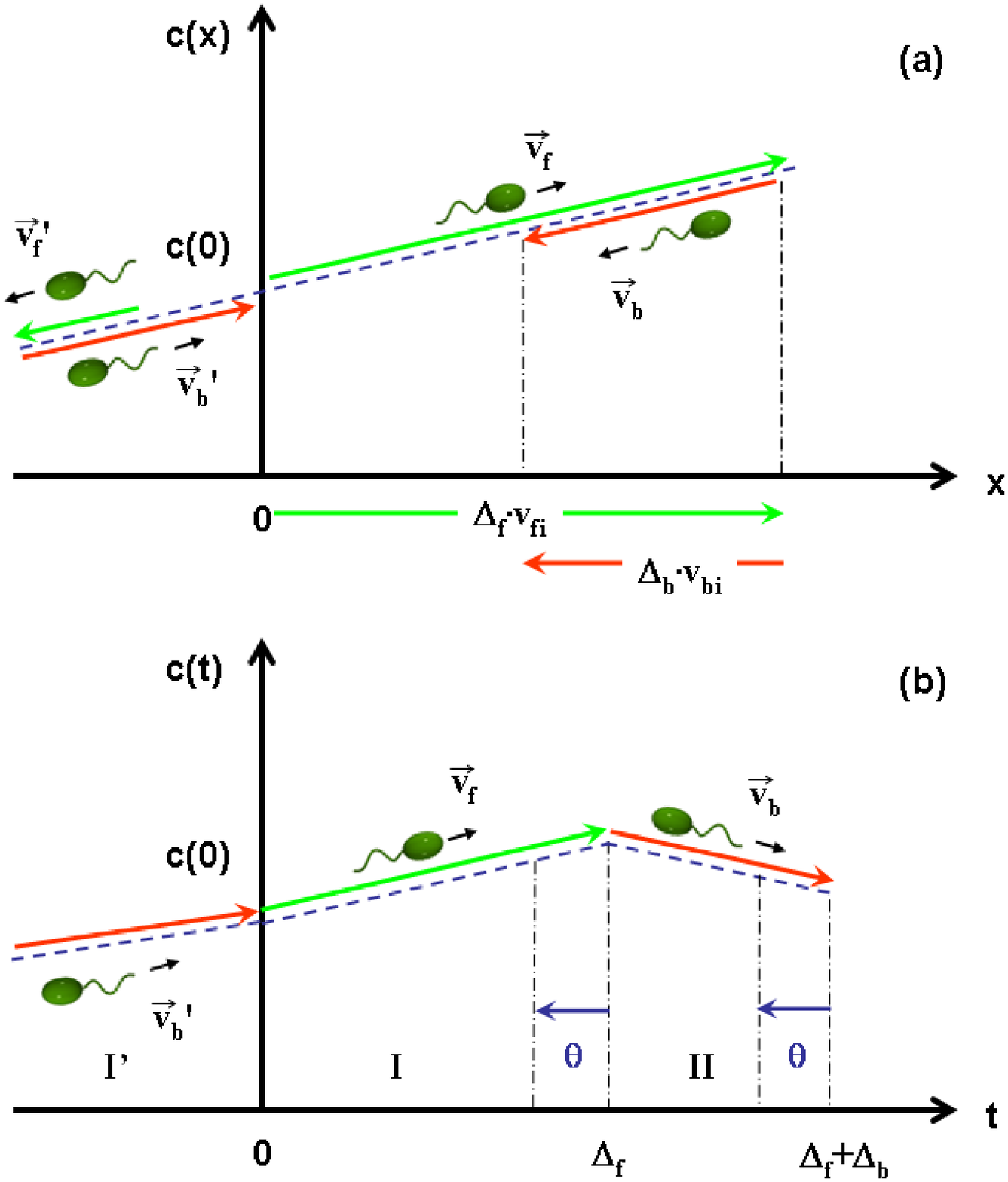}

\caption{\label{fig:LinearGradient} Migration of \emph{V. alginolyticus} in
a linear chemical gradient. (a) In the spatial domain, the chemical
gradient is specified by the dashed blue line. The green and the red
arrows correspond to forward and backward swimming segments along
the gradient. The color usage in the arrows is consistent with that
in Fig. \ref{fig:Trajectory}. $\Delta_{f}$ and $\Delta_{b}$ are
respectively the forward and backward swimming time intervals, and
$v_{fi}$ and $v_{bi}$ are respectively the forward and backward
velocity components along the chemical gradient. Note that backtracking
means $\vec{v_{b}}=-\vec{v_{f}}$. (b) The bacterial chemotactic network
processes the chemical information in the temporal domain, and the
concentration detected by the cell is depicted in (b), where $\theta$
is the memory time of the bacterium. I and II are chemosensing in
the current cycle, and I$^{\prime}$ is due to the previous cycle. }

\end{figure}

Similar to cells of \emph{E. coli}, the flagellar motor of \emph{V.
alginolyticus} has two lifetimes for the state of rotations: one ($\tau_{f}$)
for the CCW interval and one ($\tau_{b}$) for the CW interval, where
the subscripts $f$ and $b$ stand for forward and backward swimming,
respectively. \textcolor{black}{To modulate their chemotactic behaviors,
these lifetimes are affected by the local concentration of chemoeffectors
and cells' adaptation mechanism.} Unlike \emph{E. coli}, however,
CW rotation in \emph{V.} \emph{alginolyticus} causes the cell to backtrack.
Both swimming intervals are expected to depend on the ligand concentration
$c(t)$, which we assume to be chemoattractant. For small $c(t)$,
we assume that a linear response is applicable and hence,\begin{equation}
\frac{1}{\tau_{f}(t)}=\frac{1}{\tau_{f}}\left[1-\intop_{-\infty}^{t}dt'R_{f}(t-t')c(t')\right],\label{eq:forward rate}\end{equation}

\begin{equation}
\frac{1}{\tau_{b}(t)}=\frac{1}{\tau_{b}}\left[1-\intop_{-\infty}^{t}dt'R_{b}(t-t')c(t')\right],\label{eq:backward rate}\end{equation}
where $\tau_{f}$ and $\tau_{b}$ are the steady-state values, and
$R_{f}(t)$ and $R_{b}(t)$ are the memory (or the response) functions,
which are not necessarily the same for the two swimming intervals.
In the above, an exposure to the ligand causes the forward lifetime
to increase, and is consistent with our observations in \emph{V. alginolyticus}
(8). Linearity of Eqs. \ref{eq:forward rate} and \ref{eq:backward rate}
suggests that it is possible to examine one delay time $\theta$ at
a time and sum up all possible delays at the end. Following de Gennes,
we write $R_{s}(t)=\alpha_{s}\delta(t-\theta)$, where the strength
of the response $\alpha_{s}$ ($s=f,\, b$) has the dimension of volume.
Next, we consider a cell moving in a chemical gradient as depicted
in Fig. \ref{fig:LinearGradient}. Our aim is to calculate the displacement
$x_{i}$ along the gradient in one cycle, $\Delta_{f}+\Delta_{b}$,
which leads to a mean drift velocity $V=\bar{x_{i}}/(\tau_{f}+\tau_{b})$
after averaging over $\Delta_{f}$ and $\Delta_{b}$. Because a cell
randomizes its swimming direction at the end of the backward interval
by a flick, the motions in two consecutive cycles are uncorrelated.
This allows us to place the origin of time ($t=0$) at the beginning
of the forward run. \textcolor{black}{Assuming that the forward run
time is Poisson distributed, the }surviving probability of a cell
swimming forward up to $\Delta_{f}$ is given by,\begin{equation}
P_{f}(\Delta_{f})=\exp\left[-\intop_{0}^{\Delta_{f}}dt'\frac{1}{\tau_{f}(t')}\right]\simeq\exp(-\frac{\Delta_{f}}{\tau_{f}})\left[1+\frac{\alpha_{f}}{\tau_{f}}\intop_{-\theta}^{\Delta_{f}-\theta}dt'c(t')\right],\label{eq:for surv prob}\end{equation}
and the probability that it stops immediately after $\Delta_{f}$
is $-\partial P_{f}(\Delta_{f})/\partial\Delta_{f}$. Likewise, the
surviving probability of a cell swimming backwards from $\Delta_{f}$
to $\Delta_{f}+\Delta_{b}$ is given by,\begin{equation}
P_{b}(\Delta_{b},\Delta_{f})=\exp\left[-\intop_{\Delta_{f}}^{\Delta_{f}+\Delta_{b}}dt'\frac{1}{\tau_{b}(t')}\right]\simeq\exp(-\frac{\Delta_{b}}{\tau_{b}})\left[1+\frac{\alpha_{b}}{\tau_{b}}\intop_{\Delta_{f}-\theta}^{\Delta_{f}+\Delta_{b}-\theta}dt'c(t')\right],\label{eq:back surv prob}\end{equation}
and the stopping probability at the end of the backward run is $-\partial P_{b}(\Delta_{b},\Delta_{f})/\partial\Delta_{b}$.
It follows that the net mean displacement in one cycle is given by,\begin{eqnarray}
\bar{x}_{i} & \mathbf{\equiv} & \bar{x}_{fi}+\bar{x}_{bi}=\left\langle \intop_{0}^{\infty}d\Delta_{f}\left(-\frac{\partial P_{f}(\Delta_{f})}{\partial\Delta_{f}}\right)v_{fi}\Delta_{f}\right\rangle \nonumber \\
 &  & +\left\langle \intop_{0}^{\infty}d\Delta_{f}\left(-\frac{\partial P_{f}(\Delta_{f})}{\partial\Delta_{f}}\right)\intop_{0}^{\infty}d\Delta_{b}\left(-\frac{\partial P_{b}(\Delta_{b},\Delta_{f})}{\partial\Delta_{b}}\right)v_{bi}\Delta_{b}\right\rangle ,\label{eq:displacement}\end{eqnarray}
where $\bar{x}_{fi}$ and $\bar{x}_{bi}$ represent respectively the
mean displacement during the forward ($\Delta_{f}$) and the backward
($\Delta_{b}$) swimming interval, and $\langle...\rangle$ designates
the angular average for $v_{fi}$ and $v_{bi}$. For the linear gradient
\textcolor{black}{depicted} in Fig. \ref{fig:LinearGradient}, the
concentration experienced by the cell can be represented as $c(t)=c_{0}+\nabla c\cdot v_{fi}\cdot t$
for $0\le t<\Delta_{f}$ and $c(t)=c_{0}+\nabla c\cdot v_{fi}\cdot\Delta_{f}+\nabla c\cdot v_{bi}\cdot(t-\Delta_{f})$
for $\Delta_{f}\le t<\Delta_{f}+\Delta_{b}$. Since $c_{0}$ is determined
by the velocity in the previous cycle, it does not contribute to the
above integrations after angular \textcolor{black}{averaging}. Although
the calculation of Eq. \ref{eq:displacement} is tedious\textcolor{black}{,
which is given in Appendix} A, the final result is straightforward:\begin{eqnarray}
\bar{x_{i}} & = & \left\{ \alpha_{f}\tau_{f}^{2}\langle v_{fi}^{2}\rangle\exp(-\frac{\theta}{\tau_{f}})\right.\nonumber \\
 &  & \left.+\alpha_{b}\left[\frac{\tau_{f}^{2}\tau_{b}^{2}}{\tau_{f}-\tau_{b}}\langle v_{fi}v_{bi}\rangle\left(\frac{1}{\tau_{b}}\exp(-\frac{\theta}{\tau_{f}})-\frac{1}{\tau_{f}}\exp(-\frac{\theta}{\tau_{b}})\right)+\tau_{b}^{2}\langle v_{bi}^{2}\rangle\exp(-\frac{\theta}{\tau_{b}})\right]\right\} \nabla c.\label{eq:displacement1}\end{eqnarray}

The first term in the \textcolor{black}{curly} bracket of Eq. \ref{eq:displacement1}
is the displacement during the forward interval, and the second term
is the displacement during the backward interval. It is noteworthy
that during the second interval, there is a cross term proportional
to $\frac{\tau_{f}^{2}\tau_{b}^{2}}{\tau_{f}-\tau_{b}}\langle v_{fi}v_{bi}\rangle\left(\frac{1}{\tau_{b}}\exp(-\frac{\theta}{\tau_{f}})-\frac{1}{\tau_{f}}\exp(-\frac{\theta}{\tau_{b}})\right)$,
which results from the delay, i.e., even though the cell is moving
backwards, in the early episode of that interval, the cell still remembers
the concentration sensed during the previous forward swimming. This
gives rise to anti-correlation, since $\langle v_{fi}v_{bi}\rangle<0$,\textcolor{black}{{}
that}\textcolor{red}{{} }\textcolor{black}{contributes to a negative
displacement}. This important correlated motion adds richness to bacterial
chemotaxis and is what makes \emph{V. alginolyticus} behave differently
from \emph{E. coli}. We noted that Eq. \ref{eq:displacement1} yields
the result $\bar{x_{i}}=\left[\alpha_{f}\tau_{f}^{2}\langle v_{fi}^{2}\rangle+\alpha_{b}\tau_{b}\left(\tau_{b}\langle v_{bi}^{2}\rangle+\tau_{f}\langle v_{fi}v_{bi}\rangle\right)\right]\nabla c$
in the limit of no memory\textcolor{black}{,} $\theta\rightarrow0$.
\textcolor{black}{It is interesting that even when there is no memory,
the cross term survives because there is no direction randomization
after a forward run.} \textcolor{black}{Moreover, the total displacement
during the backward interval, can contribute positively or negatively
to the displacement, depending on the mean lifetimes $\tau_{f}$ and
$\tau_{b}$, and the swimming speeds $\vec{v}_{f}$ and $\vec{v}_{b}$.
We found the swimming pattern of }\textcolor{black}{\emph{V. alginolyticus}}\textcolor{black}{{}
is approximately symmetric with $\tau_{f}\simeq\tau_{b}$ and $|\vec{v}_{f}|\simeq|\vec{v}_{b}|$
(8), and hence $\bar{x}_{i}\simeq\alpha_{f}\tau_{f}^{2}\langle v_{fi}^{2}\rangle\nabla c$.}

\textcolor{black}{The corresponding quantity for }\textcolor{black}{\emph{E.
coli}}\textcolor{black}{{} is $\bar{x_{i}}=\alpha_{f}\tau_{f}^{2}\langle v_{fi}^{2}\rangle\phi_{0}\nabla c$
when $\theta\rightarrow0$, where the subscript $f$ stands for the
forward run (or CCW rotation) and $\phi_{0}=\tau_{CCW}/(\tau_{CCW}+\tau_{CW})$
is the CCW bias.}\textcolor{red}{{} }\textcolor{black}{Since near a
steady state $\phi_{0}\simeq0.5$ or $\sim0.8$ according to Refs.
(3) and (12), respectively, }\textcolor{black}{\emph{E. coli}}\textcolor{black}{{}
cells produce a smaller displacement than }\textcolor{black}{\emph{V.
alginolyticus~}}\textcolor{black}{{} within one swimming cycle if everything
else is equal.}\textcolor{red}{{} }Using $v_{fi}=-v_{bi}=v_{i}$ and
summing up all possible delays, we found from Eq. \ref{eq:displacement1}
that the mean displacement is given by\begin{eqnarray}
\bar{x_{i}} & = & \left\{ \tau_{f}^{2}\intop_{0}^{\infty}d\theta R_{f}(\theta)\exp(-\frac{\theta}{\tau_{f}})+\tau_{b}^{2}\intop_{0}^{\infty}d\theta R_{b}(\theta)\right.\nonumber \\
 &  & \left.\times\left[\exp(-\frac{\theta}{\tau_{b}})-\frac{\tau_{f}^{2}}{\tau_{f}-\tau_{b}}\left(\frac{1}{\tau_{b}}\exp(-\frac{\theta}{\tau_{f}})-\frac{1}{\tau_{f}}\exp(-\frac{\theta}{\tau_{b}})\right)\right]\right\} \langle v_{i}^{2}\rangle\nabla c.\end{eqnarray}

The average drift speed in the gradient is $V(\equiv\kappa\nabla c)=\bar{x_{i}}/(\tau_{f}+\tau_{b})$,
which allows the chemotaxis coefficient $\kappa$ to be calculated.
In \emph{E. coli}, $\kappa$ is proportional to the diffusion coefficient
$D\simeq\frac{1}{3}\phi_{0}\langle v^{2}\rangle\tau_{f}$ and one
finds $\kappa=D\intop_{0}^{\infty}R_{f}(\theta)\exp(-\frac{\theta}{\tau_{f}})d\theta$.
For an organism exhibiting the 3-step swimming pattern, the diffusivity
is given by:\begin{equation}
D=\langle v_{i}^{2}\rangle\frac{(\tau_{f}-\tau_{b})^{2}}{\tau_{f}+\tau_{b}}=\frac{1}{3}\langle v^{2}\rangle\frac{(\tau_{f}-\tau_{b})^{2}}{\tau_{f}+\tau_{b}},\label{eq:3-step diffusivity}\end{equation}
and the chemotaxis coefficient can be written as\begin{eqnarray}
\kappa & = & \frac{D}{(\tau_{f}-\tau_{b})^{2}}\left\{ \tau_{f}^{2}\intop_{0}^{\infty}d\theta R_{f}(\theta)\exp(-\frac{\theta}{\tau_{f}})\right.\nonumber \\
 &  & \left.+\tau_{b}^{2}\intop_{0}^{\infty}d\theta R_{b}(\theta)\left[\exp(-\frac{\theta}{\tau_{b}})-\frac{\tau_{f}^{2}}{\tau_{f}-\tau_{b}}\left(\frac{1}{\tau_{b}}\exp(-\frac{\theta}{\tau_{f}})-\frac{1}{\tau_{f}}\exp(-\frac{\theta}{\tau_{b}})\right)\right]\right\} .\label{eq:chemo-coefficient}\end{eqnarray}

This calculation leads to two possible scenarios (or fundamental hypotheses)
for bacterial chemotaxis: (i) independent and (ii) shared chemosensing.
In the first case, the response function\textcolor{black}{s} in the
forward and backward interval\textcolor{black}{s} are uncorrelated,
i.e., $R_{f}(\theta)$ \textcolor{black}{and $R_{b}(\theta)$ have
different functional forms}, so that the sensing system breaks the
time reversal symmetry. In order to achieve such a control, the flagellar
motor cannot only passively receive signals from the chemotaxis network
but instead the status of the motor must be made known to the chemotaxis
regulatory network. \textcolor{black}{This may be attained either
by the flagellar motor being a part of the regulatory network or by
a feedback signal via a protein that can reset the chemotactic response.
}In short, there will be a back flow of information from the motor
to the chemotaxis network in addition to the normal chemotaxis regulation.
To optimize the drifting velocity, we applied a variational principle
to Eq. \ref{eq:chemo-coefficient}, which is delineated in Appendix
B. We used the constraints that $R_{f}(\theta)$ and $R_{b}(\theta)$
\textcolor{black}{have} constant\textcolor{black}{{} variances} $\sigma_{s}^{2}/\tau_{s}$
($s=f,\, b$) (1), yielding \begin{equation}
R_{f}(\theta)\propto\frac{\sigma_{f}}{\tau_{f}}\exp(-\frac{\theta}{\tau_{f}}),\label{eq:f-response}\end{equation}
\begin{equation}
R_{b}(\theta)\propto\frac{\sigma_{b}}{\tau_{b}}\left[\exp(-\frac{\theta}{\tau_{b}})-\frac{\tau_{f}^{2}}{\tau_{f}-\tau_{b}}\left(\frac{1}{\tau_{b}}\exp(-\frac{\theta}{\tau_{f}})-\frac{1}{\tau_{f}}\exp(-\frac{\theta}{\tau_{b}})\right)\right].\label{eq:b-response}\end{equation}

\begin{figure}
\includegraphics[scale=0.5]{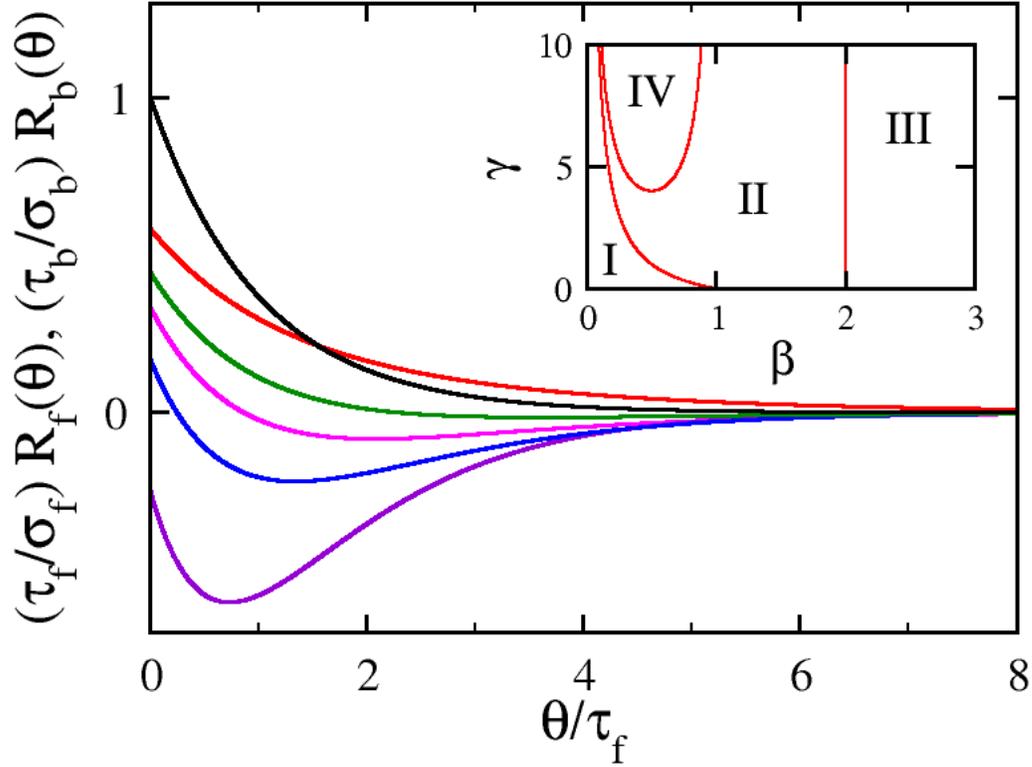}

\caption{\label{fig:schemeI} \textcolor{black}{Chemotactic Strategy I. The
bacterium uses separate response functions, $R_{f}(\theta)$ and $R_{b}(\theta)$,
for chemosensing. The figure shows the dimensionless forms of the
response functions. Here the black curve is for $R_{f}(\theta)$ and
the purple, blue, pink, green, and red curves are for $R_{b}(\theta)$
with $\beta(\equiv\tau_{b}/\tau_{f})=0.8,\,1.2,\,1.5,\,1.8$, and
2.4, respectively. The inset is the phase diagram for chemotactic
strategy II. It displays the phase boundaries between monophasic (I,
III, IV) and biphasic (II) response regimes when the chemotaxis response
obeys the relation $R(\theta)=R_{f}(\theta)=R_{b}(\theta)/\gamma$. }}

\end{figure}

\textcolor{black}{It is evident from the optimization procedure that
in order to attain the maximum possible drifting speed, the forward
response function $R_{f}(\theta)$ should be monophasic but the backward
response function $R_{b}(\theta)$ can be either monophasic or biphasic,
depending on the ratio of the two lifetimes, $\beta\equiv\tau_{b}/\tau_{_{f}}$.
Fig. \ref{fig:schemeI} displays $(\tau_{f}/\sigma_{f})R_{f}(\theta)$
and $(\tau_{b}/\sigma_{b})R_{b}(\theta)$ for different values of
$\beta=0.8,\,1.2,\,1.5,\,1.8$, and 2.4. The figure shows that the
biphasic character of $R_{b}(\theta)$ becomes more pronounced as
$\beta$ decreases towards unity but dissappears altogether for $\beta<1$,
where the response is negative for all $\theta$. An analysis shows
that the biphasic response occurs in a narrow range of $\beta$ ($1\leq\beta\leq2$),
and outside this range the response is always monophasic. This behavior
is understandable since when $\tau_{b}$ is shorter than $\tau_{f}$,
the backward interval is strongly influenced by the signal sensed
in the previous forward interval due to the memory effect. To deal
with this inconsistency between sensing and motility, the optimal
strategy is a negative monophasic response as depicted by the purple
curve ($\beta=0.8$) in Fig. \ref{fig:schemeI}. On the other hand,
when $\tau_{b}$ is longer than $\tau_{f}$, the cell would have consistent
sensing and motility so that a monophasic positive response is more
favorable, which is shown by the red curve ($\beta=2.4$) in Fig.
\ref{fig:schemeI}. In the limiting case $\tau_{b}\gg\tau_{f}$ or
$\tau_{b}\ll\tau_{f}$, Eqs. \ref{eq:chemo-coefficient}, \ref{eq:f-response},
and \ref{eq:b-response} make it clear that the chemotactic coefficient
$\kappa$ is dominated respectively by the backward or the forward
swimming interval. The situation is formally equivalent to }\textcolor{black}{\emph{E.
coli}}\textcolor{black}{{} chemotaxis, where the monophasic response
is optimal for a fast migration in a linear chemical gradient as was
concluded by de Gennes (5).}

In the case of shared chemosensing, the bacterium uses a single response
function $R(\theta)$, albeit the amplitudes of the responses may
be different in the two directions, $R_{f}(\theta)=R_{b}(\theta)/\gamma=R(\theta)$.
A simple reason for $\gamma\neq1$ could be due to different swimming
speeds $v_{f}$ and $v_{b}$, but other possibilities may also exist.
For this type of sensing, there is no breaking of time reversal symmetry
since the chemotaxis network processes information received during
the forward and the backward interval equally, and there is no need
for a back flow of information. Using Eq. \ref{eq:chemo-coefficient},
we found:\begin{eqnarray}
\kappa & = & \frac{D}{(\tau_{f}-\tau_{b})^{2}}\intop_{0}^{\infty}d\theta R(\theta)\left\{ \tau_{f}^{2}\exp(-\frac{\theta}{\tau_{f}})\right.\nonumber \\
 &  & \left.+\gamma\tau_{b}^{2}\left[\exp(-\frac{\theta}{\tau_{b}})-\frac{\tau_{f}^{2}}{\tau_{f}-\tau_{b}}\left(\frac{1}{\tau_{b}}\exp(-\frac{\theta}{\tau_{f}})-\frac{1}{\tau_{f}}\exp(-\frac{\theta}{\tau_{b}})\right)\right]\right\} .\end{eqnarray}
Applying the variational principle again (see Appendix B), we found
that the drift velocity is optimized by the following response function\begin{eqnarray}
R(\theta) & \propto & \frac{\sigma}{\tau_{f}+\tau_{b}}\left\{ \exp(-\frac{\theta}{\tau_{f}})\right.\nonumber \\
 &  & \left.+\gamma\left(\frac{\tau_{b}}{\tau_{f}}\right)^{2}\left[\exp(-\frac{\theta}{\tau_{b}})-\frac{\tau_{f}^{2}}{\tau_{f}-\tau_{b}}\left(\frac{1}{\tau_{b}}\exp(-\frac{\theta}{\tau_{f}})-\frac{1}{\tau_{f}}\exp(-\frac{\theta}{\tau_{b}})\right)\right]\right\} .\label{eq:s-response}\end{eqnarray}

\begin{figure}
\includegraphics[scale=0.5]{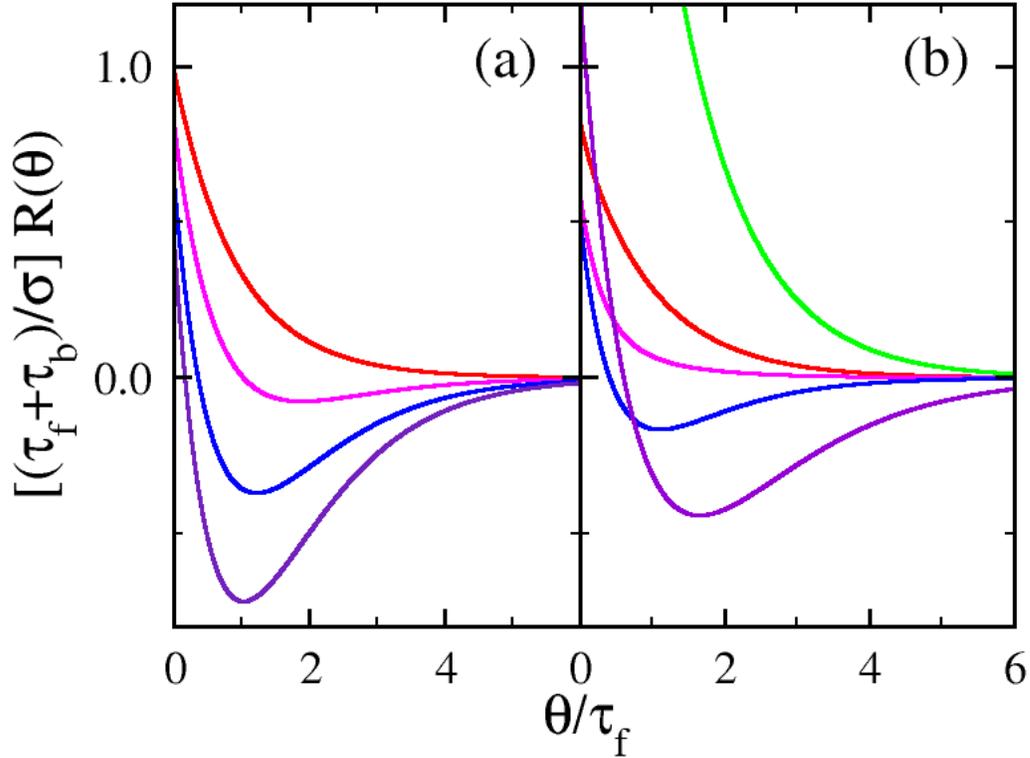}

\caption{\label{fig:schemeII} \textcolor{black}{Chemotactic Strategy II. The
bacterium shared the same response function $R(\theta)=R_{f}(\theta)=R_{b}(\theta)/\gamma$
for the forward and backward swimming intervals. In (a), $\beta=0.75$,
and $\gamma=0.1$ (red), $1.0$ (pink), $2.0$ (blue), and 3.0 (purple).
In (b), $\gamma=2$, and $\beta=0.1$ (red), $0.3$ (pink), $0.5$
(blue), 1.0 (purple), and 2.0 (green). As can be seen, for fixed $\beta\simeq0.75$,
the response becomes strongly biphasic as $\gamma$ increases. On
the other hand, for fixed $\gamma=2$, the response is monophasic
for small $\beta$, and becomes biphasic for intermediate values of
$\beta$, and returns to monophasic for $\beta\ge2$. }}

\end{figure}

\textcolor{black}{As displayed in Fig. \ref{fig:schemeII}, $R(\theta)$
can be monophasic or biphasic depending on $\gamma$ as well as the
time ratio $\beta\equiv\tau_{b}/\tau_{f}$. The biphasic regime is
bounded by $\frac{1-\beta}{\beta}\le\gamma<\frac{1}{\beta(1-\beta)}$
for $0\leq\beta<1$ and $\gamma\ge0$ for $1<\beta\leq2$, which is
displayed in the inset of Fig. \ref{fig:schemeI}. The inset shows
that the parameter space ($\gamma,\,\beta$) consists of four different
regimes with I, III, and IV being monophasic and II biphasic. Our
theory hence predicts that if a bacterium uses a single response function,
for very short ($\beta\ll1$) or very long ($\beta\gg1$) backward
swimming intervals, the biphasic response is not a good chemotactic
strategy for migration in a linear chemical gradient. The biphasic
response emerges only when $\tau_{f}$ and $\tau_{b}$ being close
(or $\beta\simeq1$), which is the case in }\textcolor{black}{\emph{V.
alginolyticus}}\textcolor{black}{{} (8). It is conspicuous that in the
limits $\beta\rightarrow1$ and $\gamma\rightarrow1$, $R(\theta)$
calculated using Eq. \ref{eq:s-response} is identical to the solution
of a critically damped harmonic oscillator, which has the interesting
property of $\int_{0}^{\infty}(1-\theta/\tau)\exp(-\theta/\tau)d\theta=0$,
i.e., the response is {}``precisely'' adaptive.}

The above two hypotheses are testable by laboratory experiments where
the bacteria are subject to a defined chemical stimulation, and one
measures the switching rate $s(t)=2/(\tau_{f}(t)+\tau_{b}(t))$ and
the forward swimming bias $\phi(t)=\tau_{f}(t)/(\tau_{f}(t)+\tau_{b}(t))$
as a function of time. For a weak stimulation, the above calculation
allows us to find,\begin{equation}
s(t)=s_{0}\left[1-\phi_{0}\intop_{-\infty}^{t}R_{f}(t-t')c(t')dt'-(1-\phi_{0})\intop_{-\infty}^{t}R_{b}(t-t')c(t')dt'\right],\label{eq:rate-full}\end{equation}
\begin{equation}
\phi(t)=\phi_{0}\left[1+(1-\phi_{0})\intop_{-\infty}^{t}\left(R_{f}(t-t')-R_{b}(t-t')\right)c(t')dt'\right],\label{eq:bias-full}\end{equation}
where $s_{0}\equiv2/(\tau_{f}+\tau_{b})$ and $\phi_{0}\equiv\tau_{f}/(\tau_{f}+\tau_{b})$
are the steady-state switching rate and the forward bias, respectively.
The expressions are significantly simplified if the perturbation is
$\delta$-in-time, $c(t)=c'\delta(t)$, and they are given by,\begin{equation}
s(t)=s_{0}\left[1-c'\left(\phi_{0}R_{f}(t)+(1-\phi_{0})R_{b}(t)\right)\right],\label{eq:rate-delta}\end{equation}
\begin{equation}
\phi(t)=\phi_{0}\left[1+c'(1-\phi_{0})\left(R_{f}(t)-R_{b}(t)\right)\right].\label{eq:bias-delta}\end{equation}
The calculation shows that if the second scenario is true and $R_{f}(t)\simeq R_{b}(t)$,
the forward bias will be weakly dependent on the time $t$, and the
switching rate is simply given by $s(t)\simeq s_{0}$$\left[1-c'R_{f}(t)\right]$.
However, if the first scenario is true, the measured $s(t)$ and $\phi(t)$
can be used to find the response function $R_{f}(t)$ and $R_{b}(t)$
using Eqs. \ref{eq:rate-delta} and \ref{eq:bias-delta}. In this
case, the following simple relations result,\begin{equation}
R_{f}(t)=\frac{1}{c'}\left[\frac{\phi(t)}{\phi_{0}}-\frac{s(t)}{s_{0}}\right],\end{equation}
\begin{equation}
R_{b}(t)=\frac{1}{c'}\left[\frac{1-\phi(t)}{1-\phi_{0}}-\frac{s(t)}{s_{0}}\right].\end{equation}

An alternative and perhaps more direct way to find $R_{f}(t)$ and
$R_{b}(t)$ is to perform conditional stimulation for individual cells.
The bacterium can be either tethered to a surface, such as in Block
et al.'s experiment (3), or freely swimming, as in Khan et al.'s experiment
(13). For tethered cells, one can apply a pulse of chemoattractant
at the moment the motor switches from CW (CCW) to CCW (CW), and record
the subsequent swimming interval $\Delta_{1f}$ ($\Delta_{1b}$),
where the subscript 1 emphasizes the interval before the first switch.
By counting the switching events up to time $t$, one can construct
a cumulative PDF (normalized by the total number of cells) $\Psi_{s}(t)$,
and the time-dependent switching rate can be obtained according to
$\tau_{s}^{-1}(t)=-\frac{d}{dt}\ln(1-\Psi_{s}(t))$, where $s=f$
or $b$. For freely swimming cells, one can use photo-active serine,
which is an attractant to \emph{V. alginolyticu}s, to stimulate cells.
If the first scenario is true, one should find that $\tau_{f}^{-1}(t)$
and $\tau_{b}^{-1}(t)$ have different time dependence or equivalently
$R_{f}(t)$ and $R_{b}(t)$ have different functional forms. However,
if the second scenario is true, there should be not much difference
between $\tau_{f}^{-1}(t)$ and $\tau_{b}^{-1}(t)$ or $R_{f}(t)\propto R_{b}(t)$.

To conclude, the 3-step motility pattern of \emph{V. alginolyticus}
discovered in our recent experiment (8) allows significant variations
\textcolor{black}{in} bacterial chemotactic behavior. These variations
can be acted on by natural selection and give rise to distinct phenotypes
observed in the wild. Compared to the 2-step swimming pattern of \emph{E.
coli}, cells of \emph{V. alginolyticus} can engage in chemosensing
and migration in both the forward and the backward swimming intervals,
and hence their {}``duty cycle'' is $\sim100\%$ as\textcolor{black}{{}
compared to $\sim50-80\%$ in }\textcolor{black}{\emph{E. coli}}\textcolor{black}{{}
(3, 12). }An important aspect in 3-step chemotaxis is backtracking
that gives those bacteria \textcolor{black}{heading down a gradient
an opportunity }to re-exploit what they find a moment earlier. In
our opinion, the full duty cycle, backtracking, and flicking are defining
characteristics of \emph{V. alginolyticus}. These significant niches
are likely selected for by the ocean environment where a quick response
to transitory signals is important. We showed that for a swimmer executing
the cyclic 3-step motility pattern, a biphasic response arises naturally
without the need to invoke cell localization as suggested for \emph{E.
coli} (1). Moreover, we showed that the biphasic response is most
effective when the forward $\tau_{f}$ and the backward $\tau_{b}$
swimming intervals are comparable. This makes biological sense since
a brief forward or a brief backward interval contributes little to
motility, and consequently a monophasic response is sufficient for
migration. This also raises the interesting question why the non-motile
CW interval in \emph{E. coli} is so long, \textcolor{black}{taking
up at least $20\%$ of the duty cycle.} If tumbling is just to change
the direction, \textcolor{black}{would }not it be better if CW interval
is shorter? An interesting possibility is that the ancestral cell
that gave {}``birth'' to \emph{E. coli} and \emph{V. alginolyticus}
was a 3-step swimmer. However, when \emph{E. coli} \textcolor{black}{became}
specilized in a different environment, which favored multiple flagella
for motility, they gave up backtracking and flicking, resulting in
a tumbly movement. In this view, then, it is not surprising that \emph{E.
coli}'s tumbling interval is long and its chemotactic response is
biphasic. 

Based on motility alone, we propose two different mechanisms, independent
and shared chemosensing, by which cells of \emph{V. alginolyticus}
can optimize their migration speed in a linear gradient. Interestingly,
the biphasic response appears in both types of chemotactic strategies.
We know very little at present how the chemotaxis network of \emph{V.
alginolyticus} regulates the 3-step motility pattern. However, based
on our calculation, it is likely that a biphasic response is also
adopted by \emph{V. alginolyticus}, and it is awaiting to be verified
in future experiments. Finally, it would be interesting to generalize
the above calculation to situations where the chemical landscape is
constantly changing, such as chemical waves (14). 

This work is supported by the NSF under the grant no. DMR-BP0646573.
\newline

\section*{Appendix A: Calculation of the Mean Displacement}

In the following we provide a more detailed derivation of the mean
displacement $\bar{x}_{i}=\bar{x}_{fi}+\bar{x}_{bi}$ in a single
3-step cycle. The displacement is made in the two time intervals,
$\Delta_{f}$ and $\Delta_{b}$, and is represented by Eq. \ref{eq:displacement}.
The concentration sensed by the bacterium is piecewise continuous
according to Fig. \ref{fig:LinearGradient} and is given by,\begin{eqnarray*}
c(t) & = & \begin{cases}
c_{0}+\nabla c\, v_{bi}'t\,, & t<0\\
c_{0}+\nabla c\, v_{fi}t\,, & 0\le t<\Delta_{f}\\
c_{0}+\nabla c\, v_{fi}\Delta_{f}+\nabla c\, v_{bi}(t-\Delta_{f})\,,\,\,\,\,\,\,\,\,\,\,\,\,\,\,\,\,\,\, & \Delta_{f}\le t<\Delta_{f}+\Delta_{b}\end{cases}\end{eqnarray*}
where the subscript $i$ designates the component of the velocity
along the gradient direction. The primed and unprimed velocities correspond
to $t<0$ (regime I$^{\prime}$) and $t\ge0$ (regimes I and II),
respectively. 

The first part of Eq. \ref{eq:displacement} is readily calculated
by integration by parts,\begin{equation}
\bar{x}_{fi}\equiv\left\langle \intop_{0}^{\infty}d\Delta_{f}\left(-\frac{\partial P_{f}(\Delta_{f})}{\partial\Delta_{f}}\right)v_{fi}\Delta_{f}\right\rangle =\left\langle \intop_{0}^{\infty}d\Delta_{f}P_{f}(\Delta_{f})v_{fi}\right\rangle \label{eq:forward_displacement}\end{equation}
where $P_{f}(\Delta_{f})$ is given by Eq. \ref{eq:for surv prob},
which contains an integration in time $t$ over the range: $-\theta\le t\le\Delta_{f}-\theta$.
Since $\Delta_{f}$ varies from $0$ to $\infty$, we have to distinguish
two cases in the integration: (i) $\Delta_{f}-\theta<0$ and (ii)
$\Delta_{f}-\theta>0$. One can deal with these two mutually exclusive
cases by the use of Heaviside functions $H(x)$, i.e., we write, 

\begin{equation}
\intop_{0}^{\infty}d\Delta_{f}(...)\equiv\intop_{0}^{\infty}d\Delta_{f}[H(\theta-\Delta_{f})+H(\Delta_{f}-\theta)](...).\end{equation}
The first Heaviside function confines the integral to $t<0$, and
since $\left\langle v_{bi}'v_{fi}\right\rangle =0$, there is no contribution
from this term. The integration constrained by the second Heaviside
function yields,\begin{equation}
\bar{x}_{fi}=\alpha_{f}\nabla c\langle v_{fi}^{2}\rangle\tau_{f}^{2}\exp\left(-\frac{\theta}{\tau_{f}}\right).\label{eq:forward-displacement}\end{equation}
This equation is identical to that found by de Gennes when he calculated
the drift velocity for \emph{E. coli} cells (5).

The second part of Eq. \ref{eq:displacement} is more complicated
because one has to take into account more possibilities. Again, we
used integration by parts to obtain,\begin{eqnarray}
\bar{x}_{bi} & \equiv & \left\langle \intop_{0}^{\infty}d\Delta_{f}\intop_{0}^{\infty}d\Delta_{b}\left(-\frac{\partial P_{f}(\Delta_{f})}{\partial\Delta_{f}}\right)\left(-\frac{\partial P_{b}(\Delta_{b},\Delta_{f})}{\partial\Delta_{b}}\right)v_{bi}\Delta_{b}\right\rangle \nonumber \\
 & = & \left\langle P_{f}(0)\intop_{0}^{\infty}d\Delta_{b}P_{b}(\Delta_{b},0)v_{bi}\right\rangle +\left\langle \intop_{0}^{\infty}d\Delta_{f}\intop_{0}^{\infty}d\Delta_{b}P_{f}(\Delta_{f})\frac{\partial P_{b}(\Delta_{b},\Delta_{f})}{\partial\Delta_{f}}v_{bi}\right\rangle .\end{eqnarray}
Let the first term in the above equation be $\bar{x}_{bi}^{\langle bb\rangle}$
and the second term be $\bar{x}_{bi}^{\langle bf\rangle}$. Since
$P_{f}(0)=1$ and $P_{b}(\Delta_{b},0)=P_{f}(\Delta_{b})$, it follows
that the integration in the first term is identical to Eq. \ref{eq:forward_displacement}
with the replacement of the subscript $f$~ by $b$. This yields,\begin{equation}
\bar{x}_{bi}^{\langle bb\rangle}=\alpha_{b}\nabla c\langle v_{bi}^{2}\rangle\tau_{b}^{2}\exp\left(-\frac{\theta}{\tau_{b}}\right).\label{eq:back-displacement}\end{equation}
Now, lets examine the anti-correlation term $\bar{x}_{bi}^{\langle bf\rangle}$,
which corresponds to the situation when the bacterium swims down the
gradient but it still keeps its {}``old good memory''. Dropping
the nonlinear terms in concentration $c$, we found,\begin{equation}
\bar{x}_{bi}^{\langle bf\rangle}=\frac{\alpha_{b}}{\tau_{b}}\left\langle \intop_{0}^{\infty}d\Delta_{f}\intop_{0}^{\infty}d\Delta_{b}\exp\left(-\frac{\Delta_{f}}{\tau_{f}}\right)\exp\left(-\frac{\Delta_{b}}{\tau_{b}}\right)\frac{\partial}{\partial\Delta_{f}}\intop_{\Delta_{f}-\theta}^{\Delta_{f}+\Delta_{b}-\theta}dt\, c(t)\, v_{bi}\right\rangle .\label{eq:anticorrelation}\end{equation}
When integrating over $\Delta_{f}$, there are two possibilities for
the lower limit of the $t$-integration, i.e. either $\Delta_{f}-\theta<0$
or $\Delta_{f}-\theta\ge0$. These will be delimited by the Heaviside
functions as before. For each of these cases, while integrating over
$\Delta_{b}$, there are additional possibilities for the upper limit
of the $t$-integration. For the first case, when $\Delta_{f}-\theta<0$,
there are three possibilities: (i) $\Delta_{f}-\theta\le\Delta_{f}+\Delta_{b}-\theta\le0$,
(ii) $0\le\Delta_{f}+\Delta_{b}-\theta\le\Delta_{f}$, and (iii) $\Delta_{f}\le\Delta_{f}+\Delta_{b}-\theta\le\Delta_{f}+\Delta_{b}$,
corresponding to the regimes I$^{\prime}$, I, and II in Fig. \ref{fig:LinearGradient},
respectively. However, since motion is uncorrelated after a flick
or $\left\langle v_{bi}'v_{bi}\right\rangle =0$, the first possibility
does not contribute to the displacement. In the second case, when
$\Delta_{f}-\theta\ge0$, there are two additional possibilities:
(iv) $\Delta_{f}-\theta\le\Delta_{f}+\Delta_{b}-\theta\le\Delta_{f}$
and (v) $\Delta_{f}\le\Delta_{f}+\Delta_{b}-\theta\le\Delta_{f}+\Delta_{b}$,
corresponding to the regimes I and II in Fig. \ref{fig:LinearGradient},
respectively. The corresponding time integrals for the above four
possibilities (ii-v) are given by\begin{eqnarray}
\intop_{\Delta_{f}-\theta}^{\Delta_{f}+\Delta_{b}-\theta}dt\, c(t) & = & \begin{cases}
c_{0}\left(\Delta_{f}+\Delta_{b}-\theta\right)+\frac{1}{2}\nabla c\, v_{fi}\left(\Delta_{f}+\Delta_{b}-\theta\right)^{2}\,, & (ii)\\
c_{0}\Delta_{f}+\frac{1}{2}\nabla c\, v_{fi}\Delta_{f}^{2}+c_{1}\left(\Delta_{b}-\theta\right)\\
\,\,\,\,\,\,\,\,\,\,\,\,\,\,\,\,\,\,\,\,\,\,\,\,\,\,\,\,\,\,+\frac{1}{2}\nabla c\, v_{bi}\left(\Delta_{b}-\theta\right)^{2}\,, & (iii)\\
c_{0}\Delta_{b}+\nabla c\, v_{fi}\Delta_{b}\left(\Delta_{f}+\frac{1}{2}\Delta_{b}-\theta\right)\,, & (iv)\\
c_{0}\theta+\nabla c\, v_{fi}\theta\left(\Delta_{f}-\frac{\theta}{2}\right)+c_{1}\left(\Delta_{b}-\theta\right)\\
\,\,\,\,\,\,\,\,\,\,\,\,\,\,\,\,\,\,\,\,\,\,\,\,\,\,\,\,\,\,\,\,\,\,+\frac{1}{2}\nabla c\, v_{bi}\left(\Delta_{b}-\theta\right)^{2}\,,\,\,\,\,\, & (v)\end{cases}\end{eqnarray}
where $c_{1}\equiv c_{0}+\nabla c\, v_{fi}\Delta_{f}$. Using the
above expressions, we take the derivative with respect to $\Delta_{f}$
to obtain,\begin{eqnarray}
\frac{\partial}{\partial\Delta_{f}}\intop_{\Delta_{f}-\theta}^{\Delta_{f}+\Delta_{b}-\theta}dt\, c(t) & = & \begin{cases}
c_{0}+\nabla c\, v_{fi}\left(\Delta_{f}+\Delta_{b}-\theta\right)\,, & (ii)\\
c_{0}+\nabla c\, v_{fi}(\Delta_{f}+\Delta_{b}-\theta)\,, & (iii)\\
\nabla c\, v_{fi}\Delta_{b}\,, & (iv)\\
\nabla c\, v_{fi}\Delta_{b}\,.\,\,\,\,\,\,\,\,\,\,\,\,\,\,\,\,\,\,\,\,\,\,\,\,\,\,\,\,\,\,\,\,\,\,\,\,\, & (v)\end{cases}\label{eq:derivatives}\end{eqnarray}
Again, using the Heaviside function to represent these four non-trivial
possibilities, we have the following identity\begin{eqnarray}
\intop_{0}^{\infty}d\Delta_{f}\intop_{0}^{\infty}d\Delta_{b}(...) & = & \intop_{0}^{\infty}d\Delta_{f}\intop_{0}^{\infty}d\Delta_{b}\nonumber \\
 &  & \left\{ H(\theta-\Delta_{f})\left[H(\Delta_{f}+\Delta_{b}-\theta)H(\theta-\Delta_{b})+H(\Delta_{b}-\theta)H(\theta)\right]\right.\nonumber \\
 &  & \left.+H(\Delta_{f}-\theta)\left[H(\Delta_{b})H(\theta-\Delta_{b})+H(\Delta_{b}-\theta)H(\theta)\right]\right\} (...).\end{eqnarray}
Substituting this equation into Eq. \ref{eq:anticorrelation}, we
found,\begin{eqnarray}
\bar{x}_{bi}^{\langle bf\rangle} & = & \frac{\alpha_{b}}{\tau_{b}}\left\langle \intop_{0}^{\infty}d\Delta_{f}\intop_{0}^{\infty}d\Delta_{b}\exp\left(-\frac{\Delta_{f}}{\tau_{f}}\right)\exp\left(-\frac{\Delta_{b}}{\tau_{b}}\right)\right.\nonumber \\
 &  & \times\left\{ H(\theta-\Delta_{f})\left[H(\Delta_{f}+\Delta_{b}-\theta)H(\theta-\Delta_{b})(ii)+H(\Delta_{b}-\theta)H(\theta)(iii)\right]\right.\nonumber \\
 &  & +\left.H(\Delta_{f}-\theta)\left[H(\Delta_{b})H(\theta-\Delta_{b})(iv)+H(\Delta_{b}-\theta)H(\theta)(v)\right]\right\} v_{bi}\Biggr\rangle,\end{eqnarray}
where $(ii)$, $(iii)$, $(iv)$, and $(v)$ are the terms given in
Eq. \ref{eq:derivatives}. The four integrations in the above equation
are delimited by different combinations of Heaviside functions, yielding
different lower and upper integration limits for each integral. Designating
these integrals as $\left(\bar{x}_{bi}^{\langle bf\rangle}\right)_{ii}$,
$\left(\bar{x}_{bi}^{\langle bf\rangle}\right)_{iii}$, $\left(\bar{x}_{bi}^{\langle bf\rangle}\right)_{iv}$,
and $\left(\bar{x}_{bi}^{\langle bf\rangle}\right)_{v}$, we found,\begin{eqnarray}
\left(\bar{x}_{bi}^{\langle bf\rangle}\right)_{ii} & = & \frac{\alpha_{b}}{\tau_{b}}\nabla c\left\langle v_{fi}v_{bi}\right\rangle \intop_{0}^{\theta}d\Delta_{f}\intop_{\theta-\Delta_{f}}^{\theta}d\Delta_{b}\exp\left(-\frac{\Delta_{f}}{\tau_{f}}\right)\exp\left(-\frac{\Delta_{b}}{\tau_{b}}\right)(\Delta_{f}+\Delta_{b}-\theta)\nonumber \\
 & = & \alpha_{b}\nabla c\left\langle v_{fi}v_{bi}\right\rangle \frac{\tau_{f}}{\tau_{f}-\tau_{b}}\exp\left[-\theta\left(\frac{1}{\tau_{b}}+\frac{1}{\tau_{f}}\right)\right]\nonumber \\
 &  & \times\left\{ \tau_{b}^{2}\left[\exp\left(\frac{\theta}{\tau_{b}}\right)-1\right]-\tau_{f}^{2}\left[\exp\left(\frac{\theta}{\tau_{f}}\right)-1\right]+\theta\left(\tau_{f}-\tau_{b}\right)\right\} ,\end{eqnarray}

\begin{eqnarray}
\left(\bar{x}_{bi}^{\langle bf\rangle}\right)_{iii} & = & \frac{\alpha_{b}}{\tau_{b}}\nabla c\left\langle v_{fi}v_{bi}\right\rangle \intop_{0}^{\theta}d\Delta_{f}\intop_{\theta}^{\infty}d\Delta_{b}\exp\left(-\frac{\Delta_{f}}{\tau_{f}}\right)\exp\left(-\frac{\Delta_{b}}{\tau_{b}}\right)(\Delta_{f}+\Delta_{b}-\theta)\nonumber \\
 & = & \alpha_{b}\nabla c\left\langle v_{fi}v_{bi}\right\rangle \tau_{f}\left[\tau_{f}+\tau_{b}-\left(\tau_{f}+\tau_{b}+\theta\right)\exp\left(-\frac{\theta}{\tau_{f}}\right)\right]\exp\left(-\frac{\theta}{\tau_{b}}\right),\end{eqnarray}

\begin{eqnarray}
\left(\bar{x}_{bi}^{\langle bf\rangle}\right)_{iv} & = & \frac{\alpha_{b}}{\tau_{b}}\nabla c\left\langle v_{fi}v_{bi}\right\rangle \intop_{\theta}^{\infty}d\Delta_{f}\intop_{0}^{\theta}d\Delta_{b}\exp\left(-\frac{\Delta_{f}}{\tau_{f}}\right)\exp\left(-\frac{\Delta_{b}}{\tau_{b}}\right)\Delta_{b}\nonumber \\
 & = & \alpha_{b}\nabla c\left\langle v_{fi}v_{bi}\right\rangle \tau_{f}\left[\tau_{b}-\left(\theta+\tau_{b}\right)\exp\left(-\frac{\theta}{\tau_{b}}\right)\right]\exp\left(-\frac{\theta}{\tau_{f}}\right),\end{eqnarray}

\begin{eqnarray}
\left(\bar{x}_{bi}^{\langle bf\rangle}\right)_{v} & = & \frac{\alpha_{b}}{\tau_{b}}\nabla c\left\langle v_{fi}v_{bi}\right\rangle \intop_{\theta}^{\infty}d\Delta_{f}\intop_{\theta}^{\infty}d\Delta_{b}\exp\left(-\frac{\Delta_{f}}{\tau_{f}}\right)\exp\left(-\frac{\Delta_{b}}{\tau_{b}}\right)\Delta_{b}\nonumber \\
 & = & \alpha_{b}\nabla c\left\langle v_{fi}v_{bi}\right\rangle \tau_{f}(\tau_{b}+\theta)\exp\left(-\frac{\theta}{\tau_{f}}\right)\exp\left(-\frac{\theta}{\tau_{b}}\right).\end{eqnarray}
In the above calculation, the terms involving $c_{0}$ do not contribute
since $\left\langle v_{bi}\right\rangle =0$. The anti-correlation
term due to all the above contributions is then given by,\begin{eqnarray}
\bar{x}_{bi}^{\langle bf\rangle} & = & \left(\bar{x}_{bi}^{\langle bf\rangle}\right)_{ii}+\left(\bar{x}_{bi}^{\langle bf\rangle}\right)_{iii}+\left(\bar{x}_{bi}^{\langle bf\rangle}\right)_{iv}+\left(\bar{x}_{bi}^{\langle bf\rangle}\right)_{v}\nonumber \\
 & = & \alpha_{b}\nabla c\left\langle v_{fi}v_{bi}\right\rangle \frac{\tau_{f}^{2}\tau_{b}^{2}}{\tau_{b}-\tau_{f}}\left[\frac{1}{\tau_{f}}\exp\left(-\frac{\theta}{\tau_{b}}\right)-\frac{1}{\tau_{b}}\exp\left(-\frac{\theta}{\tau_{f}}\right)\right].\label{eq:back-anti-displacement}\end{eqnarray}
Combining Eqs. \ref{eq:forward-displacement}, \ref{eq:back-displacement},
and \ref{eq:back-anti-displacement}, we finally obtain the mean displacement
in a given cycle for the 3-step swimmer,\begin{eqnarray}
\bar{x_{i}} & = & \alpha_{f}\nabla c\left\langle v_{fi}^{2}\right\rangle \tau_{f}^{2}\exp\left(-\frac{\theta}{\tau_{f}}\right)+\alpha_{b}\nabla c\left\langle v_{bi}^{2}\right\rangle \tau_{b}^{2}\exp\left(-\frac{\theta}{\tau_{b}}\right)\nonumber \\
 &  & +\alpha_{b}\nabla c\left\langle v_{fi}v_{bi}\right\rangle \frac{\tau_{f}^{2}\tau_{b}^{2}}{\tau_{b}-\tau_{f}}\left[\frac{1}{\tau_{f}}\exp\left(-\frac{\theta}{\tau_{b}}\right)-\frac{1}{\tau_{b}}\exp\left(-\frac{\theta}{\tau_{f}}\right)\right].\end{eqnarray}

\section*{Appendix B: Drifting Velocity Optimization}

For the first chemotactic strategy, $R_{f}(t)$ and $R_{b}(t)$ in
Eq. \ref{eq:chemo-coefficient} are independently optimized. The procedure
requires to constrain a family of response functions $R_{s}(t)$,
where $s=f,\, b$. We followed Clark and Grant's approach (1) and
assumed that $R_{s}(t)$ is finite, continuous, and decays to zero
for large $t$. The simplest way to impose the constraint is to assume
a finite variance \begin{equation}
\intop_{0}^{\infty}R_{s}^{2}(t)dt=\sigma_{s}^{2}/\tau_{s}\end{equation}
that is to be satisfied by all curves in the family. Optimizing $\kappa$
with the above constraint is equivalent to 

\begin{equation}
\frac{\delta}{\delta R_{s}(t)}\intop_{0}^{\infty}dt\left[R_{s}(t)K_{s}(t)-\lambda\left(R_{s}^{2}(t)-\frac{\sigma_{s}^{2}}{\tau_{s}}\right)\right]=0\end{equation}
where $K_{s}(t)$ is the kernel that weights the forward ($s=f$)
and the backward $(s=b)$ response functions,\begin{equation}
K_{f}(t)=\exp\left(-\frac{t}{\tau_{f}}\right),\end{equation}
\begin{equation}
K_{b}(t)=\exp(-\frac{\theta}{\tau_{b}})-\frac{\tau_{f}^{2}}{\tau_{f}-\tau_{b}}\left(\frac{1}{\tau_{b}}\exp(-\frac{\theta}{\tau_{f}})-\frac{1}{\tau_{f}}\exp(-\frac{\theta}{\tau_{b}})\right).\end{equation}
Aside from normalization constants, the optimized response functions
are given in Eqs. \ref{eq:f-response} and \ref{eq:b-response}.

The similar procedure can also be applied to the second chemotactic
strategy, resulting in the optimized response function given by Eq.
\ref{eq:s-response}.\\
\\
\\

\noindent \textbf{\Large References}\\
\textbf{\Large }\\
1. Clark, D. A., and L. C. Grant, 2005. The bacterial chemotactic
response reflects a compromise between transient and steady-state
behavior. \textit{Proc. Natl. Acad. Sci. USA} 102: 9150-9155.\\
\\
2. Celani, A., and M. Vergassola, 2010. Bacterial strategies for
chemotaxis response. \textit{Proc. Natl. Acad. Sci. USA} 107:1391-1396.\\
\\
3. Block, S. M., J. E. Segall, and H. C. Berg, 1983. Adaptation
kinetics in bacterial chemotaxis. \textit{J. Bacteriol.} 154:312-323.\\
\\
4. Schnitzer, M. J., S. M. Block, H. C. Berg, and E. M. Purcell,
1990. Strategies for chemotaxis. \textit{Symp. Soc. Gen. Microbiol.}
46:15-34.\\
\\
5. de Gennes, P., 2004. Chemotaxis: the role of internal delays.
\textit{Eur. Biophys. J.} 33:691-693.\\
\\
6. Tu, Y., T. S. Shimizu, and H. C. Berg, 2008. Modeling the chemotactic
response of \textit{Escherichia coli} to time-varying stimuli. \textit{Proc.
Natl. Acad. Sci. USA} 105: 14855-14860.\\
\\
7. Segall, J. E., S. M. Block, and H. C. Berg, 1986. Temporal comparisons
in bacterial chemotaxis. \textit{Proc. Natl. Acad. Sci. USA} 83:8987-8991.\\
\\
8. Xie, L., T. Altindal, S. Chattopadhyay, and X. L. Wu, 2010.
Bacterial flagellum as a propeller and as a rudder: new modes of bacterial
swimming and chemotaxis. Submitted to \textit{Proc. Natl. Acad. Sci.
USA}.\\
\\
9. Springer, M. S., M. F. Goy, and J. Adler, 1979. Protein methylation
in behavioural control mechanisms and in signal transduction. \textit{Nature}
280:279-284.\\
\\
10. Ciccarelli, F. D., T. Doerks, C. von Mering, C. J. Creevey,
B. Snel, and P. Bork, 2006. Toward automatic reconstruction of a highly
resolved tree of life. \textit{Science} 311:1283-1287.\\
\\
11. Kojima, M., R. Kubo, T. Yakushi, M. Homma, and I. Kawagishi,
2007. The bidirectional polar and unidirectional lateral flagellar
motors of \textit{Vibrio alginolyticus} are controlled by a single
CheY species. \textit{Mol. Microbiol.} 64:57-67.\\
\\
12. Korobkova, E., T. Emonet, J. M. G. Vilar, T. S. Shimizu, and
P. Cluzel, 2004. From molecular noise to behavioural variability in
a single bacterium. \textit{Nature} 428:574-578.\\
\\
13. Khan, S., K. Amoyaw, J. L. Spudich, G. P. Reid, and D. R. Trentham,
1992. Bacterial chemoreceptor signaling probed by flash photorelease
of a caged serine. \textit{Biophys. J.} 62:67-68.\\
\\
14. Goldstein, R. E., 1996. Traveling-wave chemotaxis. \textit{Phys.
Rev. Lett.} 77:775-778.
\end{document}